\documentclass[epj]{svjour}
\usepackage{epsfig} 
\usepackage{times}

\begin{document}
%
%
\title{Backward pion photoproduction}
\author{A.~Sibirtsev\inst{1,2}, J. Haidenbauer\inst{3,4}, 
F. Huang\inst{3}\thanks{\emph{present address:} Dept. of Physics and 
Astronomy, University of Georgia, Athens, GA 20602, U.S.A.},
S. Krewald\inst{3,4} and U.-G.~Mei{\ss}ner\inst{1,3,4}} 
\institute{
Helmholtz-Institut f\"ur Strahlen- und Kernphysik and
Bethe Center for Theoretical Physics, 
Universit\"at Bonn, D-53115 Bonn, Germany \and
Excited Baryon Analysis Center (EBAC), Thomas Jefferson National Accelerator
Facility, Newport News, Virginia 23606, USA \and
Institut f\"ur Kernphysik and J\"ulich Centre for Hadron Physics,
Forschungszentrum J\"ulich, D-52425 J\"ulich, Germany
\and Institute for Advanced Simulations,
Forschungszentrum J\"ulich, D-52425 J\"ulich, Germany}

\date{Received: date / Revised version: date}

\abstract{We present a systematic analysis of backward pion photoproduction
for the reactions $\gamma{p}{\to}\pi^0{p}$ and $\gamma{p}{\to}\pi^+{n}$.
Regge phenomenology is applied at invariant collision energies above 3~GeV
in order to fix the reaction amplitude. 
A comparision with older data on $\pi^0$- and $\pi^+$-photoproduction at
$\vartheta{=}180^\circ$ indicates that the high-energy limit as given by
the Regge calculation could be reached possibly at energies of around
$\sqrt{s}\simeq 3$~GeV. 
In the energy region of $\sqrt{s}{\le}2.5$~GeV, covered by the new
measurements of $\gamma{p}{\to}\pi^0{p}$ differential cross sections at
large angles at ELSA, JLab, and LEPS, we see no clear signal for a
convergence towards the Regge results.
The baryon trajectories obtained in our analysis are in good agreement with
those given by the spectrum of excited baryons.
}


\authorrunning{A. Sibirtsev {\it et al.} }
\titlerunning{Backward pion photoproduction}
\maketitle

\section{Introduction}
Recently, backward pion photoproduction in the $\gamma{p}{\to}\pi^0{p}$
reaction attracted significant interest at ELSA~\cite{Bartholomy},
JLab~\cite{Dugger} and LEPS~\cite{Sumihama}. New precise data were obtained
with the aim to find evidence for high-mass resonances, most of which are 
not well established~\cite{PDG}. The measurements were done at different
angles and up to photon energies that correspond to invariant collision
energy of $\sqrt{s}{\simeq}$2.5 GeV at ELSA and JLab, and
$\sqrt{s}{\simeq}$2.3~GeV at LEPS. As is evident from the publication of
the LEPS measurement~\cite{Sumihama}, their data and those collected at
ELSA are in disagreement at many of the available energies.

Naturally, the disagreement between the measurements complicates the data
evaluation in terms of a partial wave analysis (PWA). For example, the
solution of the GWU group~\cite{Arndt} does not reproduce the LEPS
\cite{Sumihama} data at the higher energies. The recent analysis of the GWU
group (FA06) presented in \cite{Dugger}, which is now extended up to
2.5~GeV and was readjusted to the new CLAS (JLab) data~\cite{Dugger}, does
not describe some of the data from ELSA, noteably at more forward angles
\cite{Arndt2}.

For pion photoproduction at backward angles there is no updated
systematic analysis. Actually, the phenomenology of backward
photoproduction was last reviewed in 1971~\cite{Berger2}. Thus, in the
present paper we want to re-examine the available high-energy data on pion
photoproduction at backward angles within the Regge approach.
We also consider the very recent measurements from ELSA~\cite{Bartholomy},
JLab~\cite{Dugger} and LEPS~\cite{Sumihama}. Thereby we want to clarify to
which extent these new data are in line with the result inferred from our
Regge fit to the high-energy data. Evidently, with increasing energy the
cross sections should approach the high-energy
limit~\cite{Hoehler,Irving,Huang}, but it is unclear  in which energy
region this will take place. 

We also explore whether the available data exhibit features that could be a
signal for high-mass resonance contributions. Indeed, the near-backward
direction is the best angular region to find such signals from the
excitation of baryons. In this region no appreciable contribution is
expected from forward dif\-fractive processes that dominate the reaction in
the $t$-channel. For example, the experimental results available for
$\pi^-p{\to}\pi^-p$ and $\pi^+p{\to}\pi^+p$ scattering at backward angles
\cite{Barger9,Crittenden} indicate a sizeable variation of the differential
cross section with energy up to $\sqrt{s}{\simeq}2.9$~GeV. This observation
might be considered as a direct illustration of the contribution from
high-mass resonances to pion-nucleon scattering. Systematic analyses of
those reactions at higher
energies~\cite{Barger9,Crittenden,Berger,Sibirtsev1} showed
that from around 3~GeV upwards the data approach the high-energy limit as
given by Regge phenomenology. It will be interesting to see whether the
situation is similar in case of pion photoproduction at backward angles. 
Finally, let us note that for reactions at backward angles the Regge
approach provides a close connection between the exchange amplitudes and
the baryon spectrum.

The paper is structured in the following way: In Sect.~2 we formulate the
reaction amplitudes. The baryon trajectories used in our analysis and their
properties are discussed in Sect.~3 and Sect.~4, respectively. The results
of the fit are presented in Sect.~5. A comparison with data on pion
photoproduction at $\vartheta$=180$^0$ and other large angles is provided
in Sects.~6 and 7. The paper ends with a brief summary.

\vfill

\section{Helicity amplitudes}
In the Regge formalism it is convenient to use the $u$-channel parity
conserving helicity amplitudes $F_\lambda^\pm(\sqrt{u},s)$ with
$\lambda{=}1, 3$ being the net photon-nucleon $u$-channel
helicity~\cite{Irving,Jacob,Gell-Mann}. Here $s$ is the invariant collision
energy squared and $u$ is the squared four-momentum transfered from the
photon to the final nucleon. The superscripts indicate the quantum number
$P{\cdot}{\cal S}$ of the baryon exchange with $P$ and ${\cal S}$ being the
parity and signature of the baryon trajectory under consideration,
respectively. For the physical particles located on the baryon trajectory
the signature factor is defined as ${\cal S}{=}(-1)^{J-1/2}$, where $J$
stands for the baryon spin. From that it is clear that the Regge
classification of the baryon trajectories that can contribute to the
reaction is given in terms of the signature ${\cal S}{=}\pm1$ and parity
$P{=}\pm1$ and, therefore, one should consider four trajectories for
nucleon- and also for Delta-isobar states. Historically these trajectories
are called $\alpha$ (for ${\cal S}{=}1$ and $P{=}1$), $\beta$ (${\cal
S}{=}1$, $P{=}{-}1$), $\gamma$ (${\cal S}{=}{-}1$, $P{=}{-}1$) and $\delta$
(${\cal S}{=}{-}1$, $P{=}1$).

The relation between the $u$-channel parity conserving helicity amplitudes
$F$ and the standard CGLN invariant amplitudes ${\cal F}$
are~\cite{Storrow2,Chew,Ball} 
\begin{eqnarray}
F_1^+(\sqrt{u},s)&=&\frac{iK_1}  {16\pi} \biggl[ 2(u-m_N^2)  {\cal F}_1 
 \nonumber  \\ &+&(\sqrt{u}+m_N)(t\sqrt{u}
-m_\pi^2m_N)  {\cal F}_2  \nonumber \\
&+&m_N(t-m_\pi^2) ({\cal F}_3+{\cal F}_4)  \nonumber \\
&+&(\sqrt{u}+m_N)(u-m_N^2) ({\cal F}_3-{\cal F}_4) \biggr], \\
F_3^+(\sqrt{u},s)&=&\frac{iK_3}  {32\pi} \left[
(\sqrt{u}+m_N){\cal F}_2+{\cal F}_3+{\cal F}_4\right], \nonumber \\
F_1^-(\sqrt{u},s)&=&F_1^+(-\sqrt{u},s),  \nonumber \\
F_3^-(\sqrt{u},s)&=&-F_3^+(-\sqrt{u},s),
\label{ampli3}
\end{eqnarray}
where $K_1$ and $K_3$ are kinematical factors given by 
\begin{eqnarray}
K_1 &=&\frac{\sqrt{(\sqrt{u}-m_N)^2-m_\pi^2}}{u} , \nonumber \\
K_3 &=&\frac{\sqrt{(\sqrt{u}{+}m_N)^2{-}m_\pi^2}
[(\sqrt{u}{-}m_N)^2{-}m_\pi^2](u{-}m_N^2)
} {\sqrt{u} u}~, \nonumber \\
\label{kinem1}
\end{eqnarray}
with $m_N (m_\pi)$  the nucleon (pion) mass. Other relations between the
amplitudes defined in various representations and the $u$-channel parity
conserving helicity amplitudes can be found in Ref.~\cite{Storrow1}.

The kinematical factors of Eq.~(\ref{kinem1}) contain a singularity at
$u{=}0$. Since the value $u{=}0$ is within the physical region such a
singularity appears in any suitable set of $u$-channel amplitudes used for
Reggeization~\cite{Paschos,Beaupre,Barger1}. The CGLN invariant amplitudes
also contain a kinematical singularity at $u{=}0$ with respect to the Ball
amplitudes~\cite{Ball},  which satisfy the Mandelstam representation and
are analytic functions of $s$, $t$ and $u$, with $t$ being the squared
four-momentum transfer from the  initial to the final nucleon.

In order to isolate this kinematical singularity the Regge amplitudes are
usually parametrized via certain residue functions that provide $u$-channel
amplitudes which vary smoothly over the value $u{=}0$. The simplest way is
to introduce modified amplitudes as in Ref.~\cite{Barger1}
\begin{eqnarray}
{\tilde F}_1^\pm (\sqrt{u},s)=\frac{F_1^\pm(\sqrt{u},s)}{K_1(\pm\sqrt{u})},
\nonumber \\
{\tilde F}_3^\pm (\sqrt{u},s)=\pm
\frac{F_3^\pm(\sqrt{u},s)}{K_3(\pm\sqrt{u})}. 
\label{singu}
\end{eqnarray}
It is clear from Eqs.~(\ref{ampli3}) and (\ref{singu}) that these modified
amplitudes satisfy the MacDowell symmetry relations~\cite{MacDowell}:
\begin{eqnarray}
{\tilde F}_1^+ (\sqrt{u},s)={\tilde F}_1^- (-\sqrt{u},s), \nonumber \\
{\tilde F}_3^+ (\sqrt{u},s)={\tilde F}_3^- (-\sqrt{u},s).
\label{symmetry1}
\end{eqnarray}
In practice these symmetry relations allow to reduce the number of
parameters used for the Reggeization of the ${\tilde F}$ amplitudes. As
will be discussed below the relations in Eq.~(\ref{symmetry1}) are
significant for the Regge classification of the baryon trajectories. In
order to satisfy the MacDowell symmetry, the Regge poles must occur in
pairs with opposite parity, with trajectories and residues related by
certain conditions listed in the following. These conditions were first
discovered by Gribov~\cite{Gribov}. Additional constraints for the
$u$-channel parity conserving helicity amplitudes are given by a threshold
relation~\cite{Jackson} that connects the helicity-$1/2$ and helicity-$3/2$
amplitudes at $u{=}m_N^2$, 
\begin{eqnarray}
{\tilde F}^\pm_1(m_N,s)= 2m_N(t-m_\pi^2)\, {\tilde F}^\pm_3(m_N,s)\, .
\label{threshold}
\end{eqnarray}

Similar to the particle-exchange Feynman diagram, each amplitude $F$ is
factorized in terms of a propagator and a vertex function
as~\cite{Gell-Mann,Beaupre,Barger1}
\begin{eqnarray}
{\tilde F}^\pm_1 &=& i \gamma^\pm_1(\sqrt{u}) \, G(\alpha^\pm(u),\nu) , 
\nonumber \\
{\tilde F}^\pm_3 &=& i \gamma^\pm_3(\sqrt{u})  \, 
\frac{\alpha^\pm (\sqrt{u})-1/2}{\nu} \, G[\alpha^\pm(u),\nu],
\end{eqnarray}
where the $\gamma$'s are residue functions that satisfy the symmetry relations of
Eq.~(\ref{symmetry1}):
\begin{eqnarray}
\gamma^+_1(\sqrt{u})=\gamma^-_1(-\sqrt{u}) \,\,\,\,\, {\rm and} \,\,\,\,\,
\gamma^+_3(\sqrt{u})=\gamma^-_3(-\sqrt{u}) .
\end{eqnarray}
The invariant variable $\nu$ is defined by 
\begin{eqnarray}
\nu = \frac{s-t}{4m_N}=\frac{2s+u-2m_N^2-m_\pi^2}{4m_N}
\end{eqnarray}
and the Regge propagator is given by
\begin{eqnarray}
G[\alpha,\nu] =\frac{1{+}{\cal S}\exp[-i \pi (\alpha
{-}1/2)]}{\Gamma [\alpha {+}1/2] \, \cos[\pi \alpha]}\,\,
\left[\frac{\nu}{\nu_0}\right]^{\alpha-1/2},
\label{prop}
\end{eqnarray}
with $\nu_0{=}1$~GeV. The signature factor ${\cal S}$ was already defined
previously and $\alpha$ is the Regge baryon trajectory that satisfies also
the MacDowell symmetry,
\begin{eqnarray}
\alpha^+(\sqrt{u})=\alpha^-(-\sqrt{u}),
\label{sym2}
\end{eqnarray}
and was taken as a linear function of $u$,
\begin{eqnarray}
\alpha(\sqrt{u}) =\alpha_0 + 0.9u,
\label{traj1}
\end{eqnarray}
with an intercept $\alpha_0$ that is a free parameter. However, as will be
illustrated in the next Section, the intercept of the Regge trajectories
can also be very well constrained by the classification of the spectrum of
excited baryons.

The threshold relation Eq.~(\ref{threshold}) is explicitly imposed by
\begin{eqnarray}
\gamma^\pm_1 ={-}4m_N^2(\alpha^\pm{-}1/2)\gamma^\pm_3
{-}(u{-}m_N^2)\gamma^\pm_0,
\end{eqnarray}
where $\gamma_0$ and $\gamma_3$ are residue functions, whose
parametrizations are given below. Here the $\sqrt{u}$-dependence of the
$\gamma$ functions was dropped for simplicity. The residue functions
$\gamma_0$ and $\gamma_3$ as well as the intercept of the baryon
trajectories were fitted to the data. We use the following isospin
decomposition of the scattering amplitudes for the different charge states:
\begin{eqnarray}
\frac{1}{3}\, [N_V-\sqrt{3}N_S+2\Delta ] \,\,\,\, {\rm for} \,\,\,\, \gamma p
\to p \pi^0, \nonumber \\
 \frac{\sqrt{2}}{3}\, [N_V+\sqrt{3}N_S-\Delta ] \,\,\,\, {\rm for} \,\,\,\, 
\gamma p \to n \pi^+, \nonumber \\
 \frac{1}{3}\, [N_V+\sqrt{3}N_S+2\Delta ] \,\,\,\, {\rm
for} \,\,\,\, \gamma n \to n \pi^0, \nonumber \\
 \frac{\sqrt{2}}{3} \, [N_V-\sqrt{3}N_S-\Delta ] \,\,\,\, {\rm for} \,\,\,\,
\gamma n \to p \pi^- .
\label{isospin}
\end{eqnarray}
Here $S$ and $V$ are the isoscalar and isovector components of the
electromagnetic coupling, respectively. The symbols $N$ and $\Delta$ denote
$I{=}1/2$ and $I{=}3/2$ exchange amplitudes. The $N_S/N_V$ ratio was taken
as a constant and, finally, was fitted to the data. The parity-conserving
$u$-channel amplitudes $F$ were trans\-formed to the standard $s$-channel
helicity amplitudes, $H$, using the matrix relations~\cite{Storrow1}. The
observables are given in terms of the amplitudes $H$ in
Refs.~\cite{Barker,Benmerrouche}.

\section{Baryon trajectories}
So far we have not discussed what baryon trajectories should be included in
the analysis of backward pion photoproduction and we have not specified the
functional form of the residue functions. Actually, the functional form of
$\gamma_0(\sqrt{u})$ and $\gamma_3(\sqrt{u})$ can be fixed by the
corresponding baryon trajectory to some extent, as will be discussed in
the following.

\begin{figure}[t]
\vspace*{-5.4mm}
\centerline{\hspace*{2mm}\psfig{file=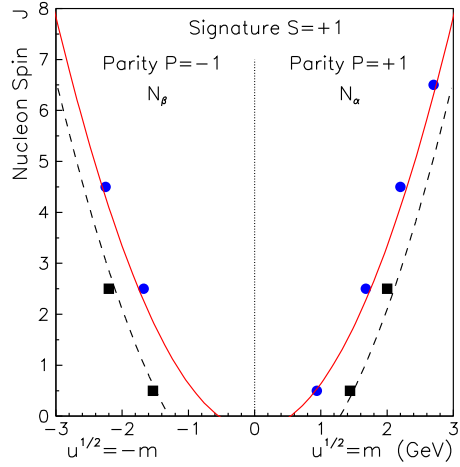,width=7.cm}}
\vspace*{-3mm}
\caption{Chew-Frautschi plot for the  $N_\alpha$ and $N_\beta$ nucleon
trajectories indicating the baryon spin $J$ as a function of the mass
($\sqrt{u}$). The results are for the signature ${\cal S}{=}{+}1$ and for
the parities $P {=}{\pm}1$. The circles represent the excited nucleon
states listed by the PDG~\cite{PDG} that belong to the leading
trajectories. The squares are states that lie on the next-to-leading or
so-called daughter trajectories. The solid line indicates the leading
trajectories $\alpha^+(\sqrt{u})$ and $\alpha^-(-\sqrt{u})$ obtained with
$\alpha_0{=}{-}0.26$, while the dashed line corresponds to the daugther
trajectories with $\alpha_0{=}{-}1.5$}
\label{phob6a}
\end{figure}

Let us first consider the nucleon trajectory with spin $J{=}1/2$ and with
$N(938)$ as the lowest lying state and thus with signature ${\cal
S}{=}{+}1$ and parity $P{=}{+}1$. Historically this trajectory is called
the $N_\alpha$-trajectory. Its contribution to the $F^+$ amplitudes should
be parametrized in terms of the $\gamma^+_0$ and $\gamma^+_3$ residues
functions and the $\alpha^+$ trajectory. The masses of the lowest excited
nucleon states with the same quantum numbers $P$ and ${\cal S}$, but with
higher spins, are listed by the PDG~\cite{PDG} as $F_{15}(1680)$,
$H_{19}(2200)$ and $K_{1,13}(2700)$, corresponding to
$J{=}\frac{5}{2},\frac{9}{2}$ and $\frac{13}{2}$, respectively.

Fig.~\ref{phob6a} shows these resonance states in the Chew-Frautschi
plot~\cite{Barger4,Fiore} of the spin $J$ as a function of $\sqrt{u}{=}m$,
so that $m$ concurs with the masses of the nucleon states listed above at
the corresponding $J$ values. The solid line is the Regge trajectory
$\alpha^+(\sqrt{u})$ given by Eq.~(\ref{traj1}) with $\alpha_0{=}{-}0.26$.
Here the intercept was simply adjusted by hand but not precisely fitted to
the baryon spectrum. Note that in this case $P{\cdot}{\cal S}{=}{+1}$.

The MacDowell symmetry implies the existence of states with the same
signature ${\cal S}{=}{+1}$ but opposite parity $P{=}{-}1$. Indeed for such
states $P{\cdot}{\cal S}{=}{-1}$ and one can apply Eq.~(\ref{sym2}). The
PDG~\cite{PDG} lists the corresponding excited states $D_{15}(1675)$ and
$G_{19}(2250)$ for spins $J{=}\frac{5}{2}$ and $\frac{9}{2}$, respectively.
These excited nucleon states are shown in Fig.~\ref{phob6a} for
${-}\sqrt{u}{=}m$ together with the $\alpha^-(-\sqrt{u})$ trajectory.
Historically this class of resonances is called $N_\beta$-trajectory.

Fig.~\ref{phob6a} illustrates that, indeed, the symmetry is valid for the
existing nucleon resonances with ${\cal S}{=}{+1}$. But it does not give
any explanation why there is no parity partner with $\frac{1}{2}^-$, i.e.
for the nucleon ground state $\frac{1}{2}^+$. Still, this fact must be
taken into account in the parameterization of the residue
functions~\cite{Storrow2,Gribov,Minkowski,Carlitz}. Thus, the residues of
the trajectory are parametrized as
\begin{eqnarray}
\gamma_0^+(\sqrt{u}) = I^\alpha \beta_0^\alpha  \left(1+\frac{\sqrt{u}}{0.938}
\right), \nonumber \\
\gamma_3^+(\sqrt{u}) = I^\alpha \frac{\beta_3^\alpha}{m_N^2-u} 
\left(1+\frac{\sqrt{u}}{0.938}
\right),
\label{resalpha}
\end{eqnarray}
where $I^\alpha$ is an isospin factor, which defines the contribution to
the different reaction channels for photoproduction of the $\pi^0$, $\pi^+$
and $\pi^-$-mesons. The constants $\beta_0^\alpha$ and $\beta_3^\alpha$
should be fixed by a fit to the data. Note that the residue of
Eq.~(\ref{resalpha}) vanishes for the $\frac{1}{2}^-$ state, i.e. at
$\sqrt{u}{=}-m_N$, so that the trajectory does not pass through an
unobserved state. 

\begin{figure}[t]
\vspace*{-3mm}
\centerline{\hspace*{2mm}\psfig{file=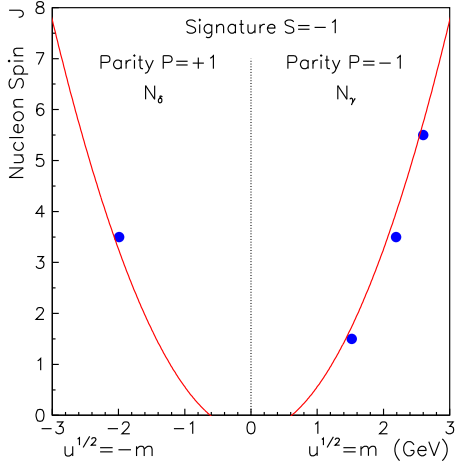,width=7.cm}}
\vspace*{-3mm}
\caption{Chew-Frautschi plot for the  $N_\gamma$ and $N_\delta$ nucleon
trajectories indicating the spin baryon $J$ as a function of the the mass ($\sqrt{u}$). 
The results are for the signature ${\cal S}{=}{-}1$ and for the 
parities $P {=}{\pm}1$. 
The circles represent the excited nucleon states listed by PDG~\cite{PDG}. 
The line indicates the trajectories $\alpha^+(\sqrt{u})$ and $\alpha^-(-\sqrt{u})$ 
for parities $P {=}{-}1$ and $P {=}{+}1$, respectively, obtained 
with $\alpha_0{=}{-}0.34$.
}
\label{phob6b}
\end{figure}

The other nucleon states with ${\cal S}{=}{+}1$ and $P{=}{\pm}1$ but with
higher masses are classified as daughter or granddaughter ({\it etc.})
trajectories. The squares in the Fig.~\ref{phob6a} show the states $P_{11}(1440)$
and $F_{15}(2000)$ for positive parity and spin $J{=}\frac{1}{2}$ and
$J{=}\frac{5}{2}$, respectively. The negative parity resonance states in this
case are $S_{11}(1535)$ and $D_{15}(2200)$ for $J{=}\frac{1}{2}$ and
$\frac{5}{2}$, respectively. The dashed line in Fig.~\ref{phob6a} is the 
linear (next-to-leading) Regge trajectory with $\alpha_0{=}{-}1.5$. 

The secondary (daugther) trajectory is
low-lying with respect to the leading trajectory (solid line).
Because of the large negative value of $\alpha_0$ its contribution to
the reaction amplitude is expected to be suppressed due to the $\nu$-dependence 
of the Regge propagator, cf. Eq.~(\ref{prop}). That is why the contributions from
secondary trajectories are frequently neglected in Regge analyses. 
But one should keep in mind that there is no solid argument in favor of 
such an expectation, because the residue constants $\beta_0^\alpha$ and
$\beta_3^\alpha$ are free (phenomenological) parameters and could be large.

\begin{figure}[t]
\vspace*{-3mm}
\centerline{\hspace*{2mm}\psfig{file=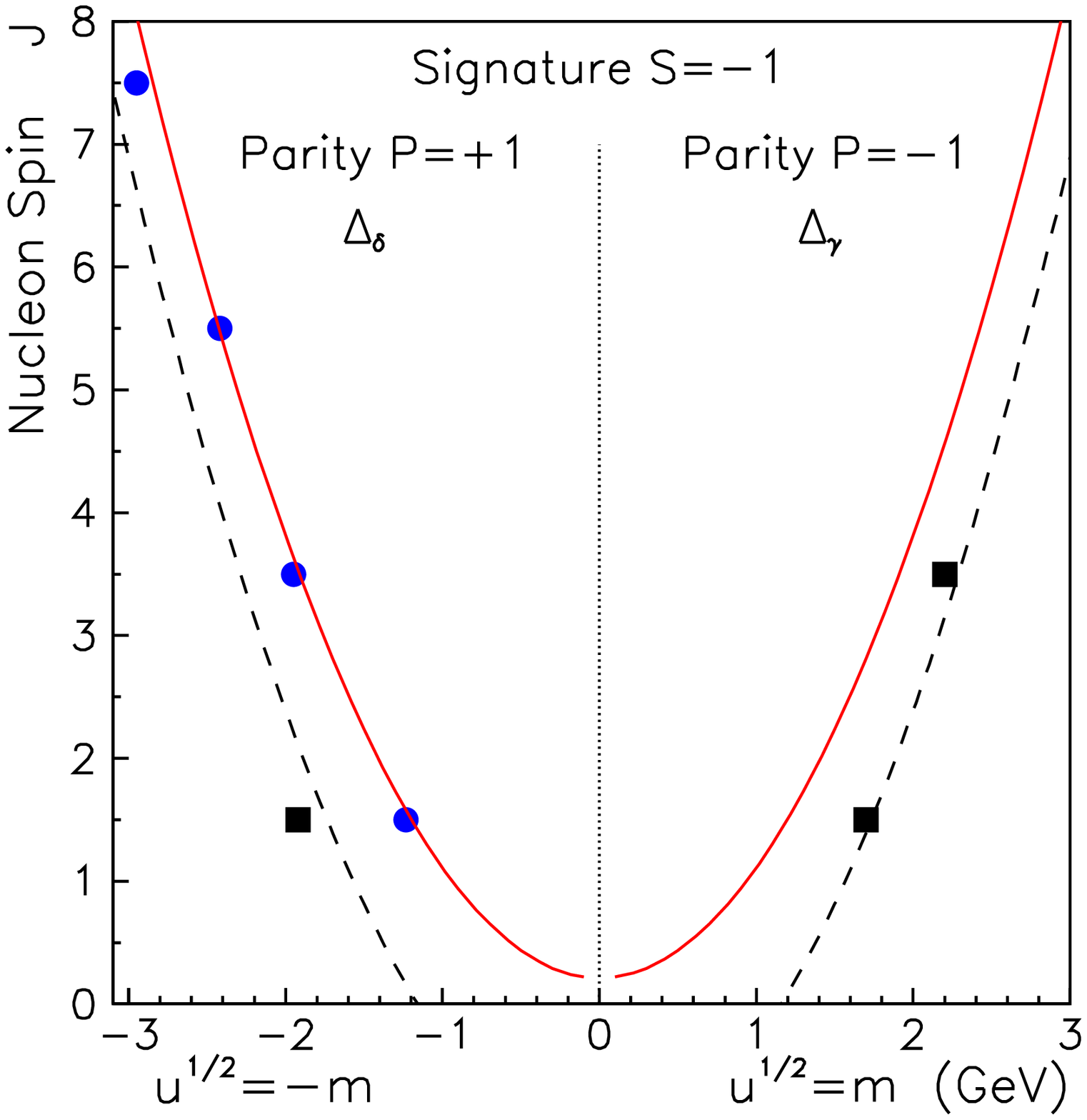,width=7.cm}}
\vspace*{-3mm}
\caption{Chew-Frautschi plot for the $\Delta_\gamma$ and $\Delta_\delta$
trajectories indicating the spin baryon $J$ as a function of the mass ($\sqrt{u}$). 
The results are for the signature ${\cal S}{=}{-}1$ and for the 
parities $P {=}{\pm}1$. 
The circles represent excited $\Delta$ states listed by the PDG~\cite{PDG} 
that belong to the leading trajectory.
The solid line indicates the leading trajectories $\alpha^+(\sqrt{u})$
and $\alpha^-(-\sqrt{u})$ for parities $P{=}{-}1$ and $P
{=}{+}1$, respectively based on the intercept $\alpha_0{=}$0.21. 
The squares are states that lie on the next-to-leading trajectory. 
The dashed line is the secondary trajectory given by Eq.~(\ref{traj1}) 
with $\alpha_0{=}{-}$1.21.
}
\label{phob6c}
\end{figure}

Now let us consider the nucleon states with signature factor ${\cal
S}{=}{-}1$. For parity $P{=}{-1}$ the PDG~\cite{PDG} lists the following
resonances: $D_{13}(1520)$, $G_{17}(2190)$ and $I_{1,11}(2600)$ with spin
$J{=}\frac{3}{2}$, $\frac{7}{2}$ and $\frac{11}{2}$, respectively. These excited
nucleons belong to the so-called $N_\gamma$-trajectory and are shown in 
Fig.~\ref{phob6b}. 
The line represents the $\alpha^+(\sqrt{u})$ trajectory with 
intercept $\alpha_0{=}{-}0.34$.  

For the $N_\delta$-trajectory with ${\cal S}{=}{-}1$ and parity $P{=}{+1}$
the PDG lists $F_{17}(1990)$ with spin $\frac{7}{2}$.
There is no indication for a positive-parity partner for the $D_{13}(1520)$
resonance (which has negative parity). This is taken into account in the 
residue functions for the $N_\gamma$-trajectory, which are related to the
residues of the $N_\delta$-trajectory by the MacDowell symmetry. Namely, 
we parametrize the residues by
\begin{eqnarray}
\gamma^+_i (\sqrt{u}) =I^\gamma \beta^\gamma_i
\left(1+\frac{\sqrt{u}}{1.52}\right),
\end{eqnarray}
for $i{=}0,3$ with the constants $\beta^\gamma_0$ and $\beta^\gamma_3$
as free parameters. 

There are only a few known nucleon states that 
might be discussed in terms of the secondary trajectories
with negative signature. Therefore we do not address this question here.

We proceed further with the spectrum of the $\Delta$-resonances and the relevant
trajectories. For signature ${\cal S}{=}{-}1$ and  positive parity $P{=}{+}1$
the PDG~\cite{PDG} lists the following baryonic resonances: $P_{33}(1232)$,
$F_{37}(1950)$, $H_{3,11}(2420)$, $K_{3,15}(2950)$ with spin
$J{=}\frac{3}{2}$, $\frac{7}{2}$, $\frac{11}{2}$ and $\frac{15}{2}$, respectively. 
These resonance states form the so-called $\Delta_\delta$-trajectory and are shown in
Fig.~\ref{phob6c} by circles. The solid line is the trajectory
$\alpha^-({-}\sqrt{u})$ with $\alpha_0{=}0.21$. 
Unfortunately, there are no experimentally well identified 
$\Delta$-resonances with ${\cal S}{=}{-}1$ and $P{=}{-}1$, which would belong 
to the $\Delta_\gamma$-trajectory and would be the parity partners of the 
$\Delta_\delta$ excited baryons.

So far the partial wave analyses have not found any 
indication for the $\Delta_\gamma$ with a mass around 1.232~GeV. Therefore, we
parametrize the residue function of the $\Delta_\delta$-trajectory by
\begin{eqnarray}
\gamma^-_i (-\sqrt{u}) =I^\delta \beta^\delta_i
\left(1-\frac{\sqrt{u}}{1.232}\right),
\end{eqnarray}
for $i{=}0,3$ and with the constants $\beta^\delta_0$ and $\beta^\delta_3$
as free parameters. Note that this parameterization eliminates the 
lowest-lying $\Delta_\gamma$ state, although it is not clear whether the
other high-mass baryons of this trajectory indeed exist in nature.

The squares in Fig.~\ref{phob6c} show the states
lying on next-to-leading trajectory. The dashed lines is the secondary
trajectory given by Eq.~(\ref{traj1}) with intercept $\alpha_0{=}{-}$1.21. 

\begin{figure}[t]
\vspace*{-3mm}
\centerline{\hspace*{2mm}\psfig{file=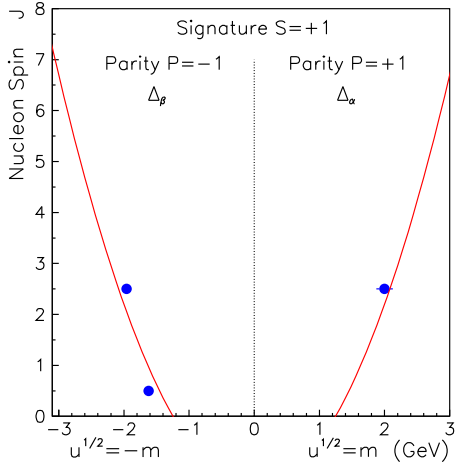,width=7.cm}}
\vspace*{-3mm}
\caption{Chew-Frautschi plot for the $\Delta_\alpha$ and $\Delta_\beta$
trajectories indicating the baryon spin $J$ as a function of the mass ($\sqrt{u}$). 
The results are for the signature ${\cal S}{=}{+}1$ and for the 
parities $P {=}{\pm}1$. 
The circles represent the excited states listed by the PDG~\cite{PDG} that
belong to the leading trajectory. 
The solid lines are the trajectories $\alpha^+(\sqrt{u})$
and $\alpha^-(-\sqrt{u})$ for parities $P{=}{-}1$ and $P
{=}{+}1$, respectively, with intercept $\alpha_0{=}{-}$1.4.
}
\label{phob6d}
\end{figure}

Finally we complete this short review of the nucleon trajectories with the
$\Delta$-resonances that have positive signature ${\cal S}{=}{+}1$ . For
$P{=}{-}1$ we consider the $S_{31}(1620)$ and $D_{35}(1930)$ states with
spin $J{=}\frac{1}{2}$ and $\frac{5}{2}$, respectively. The Regge trajectory
containing these resonances is called $\Delta_\beta$-trajectory. For the parity
$P{=}{+}1$ the PDG~\cite{PDG} lists one relevant resonance, the $F_{35}(2000)$,
which belongs to the $\Delta_\alpha$-trajectory. These excited baryons are
shown in the Fig.~\ref{phob6d} by circles. The lines are the Regge trajectories 
with $\alpha_0{=}{-}1.4$. 

Because of the large negative intercept of the $\Delta_\alpha$ and
$\Delta_\beta$ trajectories, their contribution to the total reaction amplitude
is substantially suppressed due to the $\nu$-dependence of the Regge propagator,
cf. Eq.~(\ref{prop}). That is the reason why in many analyses these trajectories 
were not considered.

\section{Propagator properties}
The baryon trajectories roughly determined from the nucleon and
$\Delta$-resonance spectra allow for some further constraints on their 
respective contributions to the reaction amplitude. Let us consider the 
squared Regge propagator in Eq.~(\ref{prop}) and take into account that 
\begin{eqnarray}
\Gamma\left[z+\frac{1}{2}\right]\cos[\pi z]= \pi
\Gamma^{-1}\left[\frac{1}{2}-z\right].
\end{eqnarray}
For the trajectories with signature factor ${\cal S}{=}{\pm}1$  the squared
propagator is
\begin{eqnarray}
\nonumber 
|G|^2 {=}\frac{\Gamma^2}{\pi^2} \!\!\left[\frac{1}{2}{-}\alpha(\sqrt{u})\right]\!\!
\left[2{+}2{\cal
S}\!\cos\bigl[\pi(\alpha(\sqrt{u}){-}\frac{1}{2})\bigr]\right]
\!\!\left[\frac{\nu}{\nu_0}\right]^{
2\alpha{-}1}\!\!\!\!\!\!\!\!\!. \,\, \\
\label{square}
\end{eqnarray}

\begin{figure}[t]
\vspace*{-1.5mm}
\centerline{\hspace*{2mm}\psfig{file=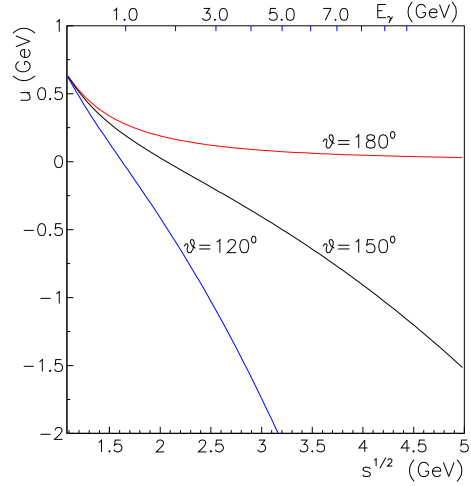,width=7cm}}
\vspace*{-3mm}
\caption{Four-momentum transfer squared $u$ as a function of the invariant 
collision energy $\sqrt{s}$ (lower axis) or photon energy (upper axis) 
for different pion production angles $\vartheta$ (in the $s$-channel).
}
\label{phob8a}
\end{figure}

An interesting feature of the Regge propagator are the zeros of
Eq.~(\ref{square}). The zeros at $u{\ge}0.6$~GeV$^2$ have no physical meaning
because they are located outside of the scattering (physical) region. This is illustrated 
in Fig.~\ref{phob8a} which shows the correspondence between the scattering
angle $\vartheta$ and the four-momentum transfer squared $u$  for different
invariant collision energies $\sqrt{s}$ or photon energies $E_\gamma$. 
Here we define the
scattering angle in the $s$-channel, {\it i.e.} as the pion production 
angle with respect to the photon beam direction given in the center-of-mass
(cm) system. The maximal value of $u$ corresponds to pion photoproduction at 
the angle $\vartheta{=}180^\circ$. 

\begin{table}[t]
\begin{center}
\caption{Leading Regge baryon trajectories with signature factor ${\cal S}$ and
parity $P$. The listed intercepts $\alpha_0$ are determined from the spectrum 
of the excited baryons. The zeros of the squared Regge propagator as given in 
Eq.~(\ref{square}), i.e. the values where $|G(u)|^2{=}0$, are listed, too. 
}
\label{summary1}
\begin{tabular}{|c|r|r|r|r|r|r|}
\hline
 & ${\cal S}$ & $P$ & $\alpha_0$ &  
\multicolumn{3}{|c|}{Zeros } \\ 
 & & &  & \multicolumn{3}{|c|}{ u (GeV$^2$) } \\ 
\hline
$N_\alpha$      & $+1$ & $+1$ & $-0.26$  & $-2.49$ & $-0.26$ & $1.96$ \\
$N_\gamma$      & $-1$ & $-1$ & $-0.34$  & $-3.51$ & $-1.85$ & $0.93$ \\
$\Delta_\delta$ & $-1$ & $+1$ & $0.21$  & $-4.12$ & $-1.9$ & $0.32$ \\
$\Delta_\beta$  & $+1$ & $-1$ & $-1.4$  & $-3.4$ & $-1.2$ & $1.0$ \\
\hline
\end{tabular}
\end{center}
\end{table}

Fig.~\ref{phob8a} makes clear that the squared four-momentum $u$ can be
positive as well as negative and thus one indeed has to care about the
kinematical singularities at the $u{=}0$ line. The zeros of the propagator of
Eq.~(\ref{prop}) at negative four-momentum transfer squared are accessible and
might be reflected in scattering observables.

The zeros and the intercepts of the various trajectories are summarized in
Table~\ref{summary1}. 
The $N_\alpha$-trajectory has its first zero at $u{=}{-}0.26$~GeV$^2$. 
Indeed, the differential cross sections of the $\pi{N}{\to}\pi{N}$ 
reaction exhibit a dip in that $u$ region, though around 
$u{\simeq}{-}0.15$~GeV$^2$~\cite{Berger,Barger3,Storrow3,Winbow}, which does 
not agree that well with the value of the trajectory.
 
The pion photoproduction data on differential cross sections show no dip at
$|u|{<}1$~GeV$^2$. This means that other trajectories than $N_\alpha$ and
$\Delta_\beta$ must dominate the reaction. The $\Delta_\delta$ trajectory 
alone cannot dominate the reaction either because then 
the cross-section ratio of $\gamma{p}{\to}\pi^0p$ to $\gamma{p}{\to}\pi^+n$ 
should be equal to two at small $|u|$. The experimental data indicate that 
this ratio is close to one at high energies. This also excludes that 
one can achieve a description of backward pion photoproduction by 
considering only the $N_\alpha$ and $\Delta_\delta$ trajectories. 
Consequently, one has to take into acount also the contribution 
from the $N_\gamma$-trajectory in the Regge analysis of backward
scattering.

\begin{figure}[t]
\vspace*{-6mm}
\centerline{\hspace*{2mm}\psfig{file=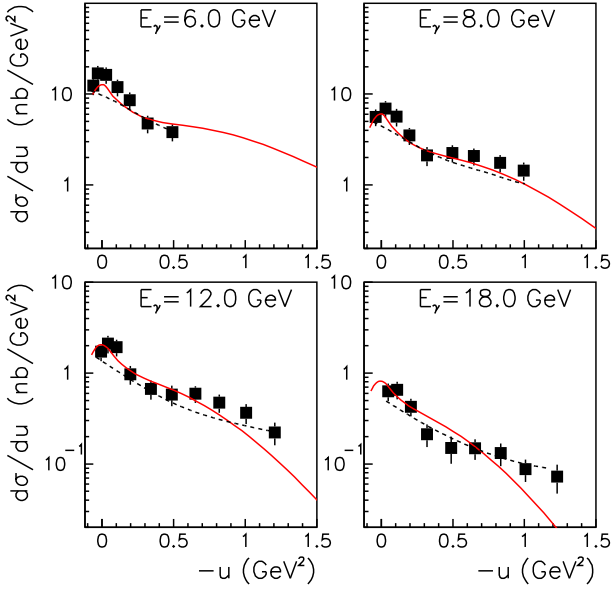,width=9.cm}}
\vspace*{-5mm}
\caption{Differential cross section for $\gamma{p}{\to}\pi^0{p}$ as a function
of the four-momentum transfer squared $u$ at different photon energies $E_\gamma$. 
The squares are SLAC data from Ref.~\cite{Tompkins}. The solid lines are
the results of our Regge model employing the parameters listed in
Table~\ref{tabp}. The dashed lines are the calculations from
Ref.~\cite{Storrow1}.}
\label{phob1}
\end{figure}

It is interesting to note that both $N_\gamma$ and $\Delta_\delta$ trajectories
have their first zeros for negative $u$-values around $u{\simeq}{-}1.9$~GeV$^2$. 
However, the available data are not precise enough to see whether this 
feature of the Regge propagator is actually reflected in the observables.
 
\section{Results at high energies}
\begin{figure}[t]
\vspace*{-6mm}
\centerline{\hspace*{2mm}\psfig{file=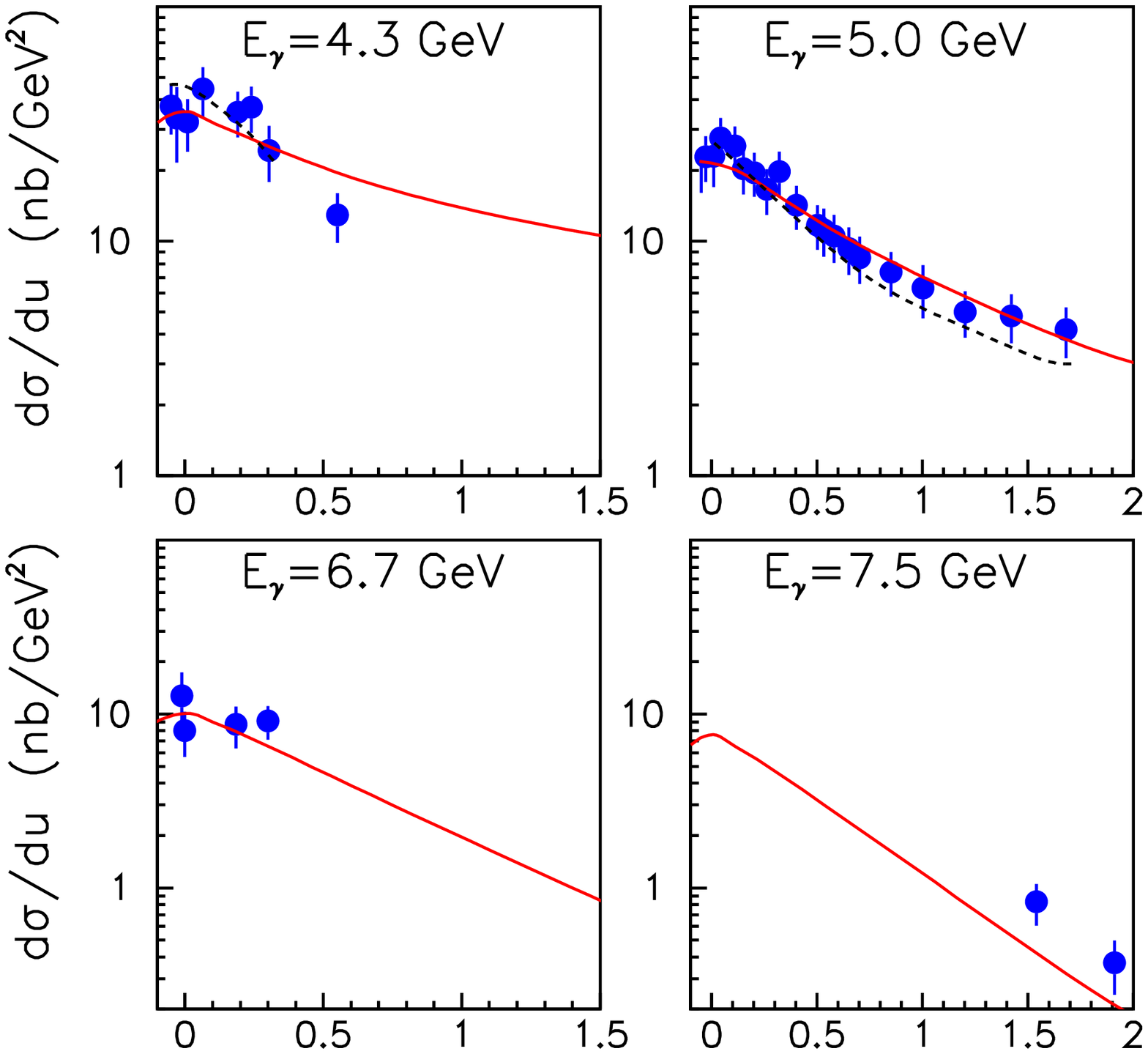,width=9.cm}}
\vspace*{-16mm}
\centerline{\hspace*{2mm}\psfig{file=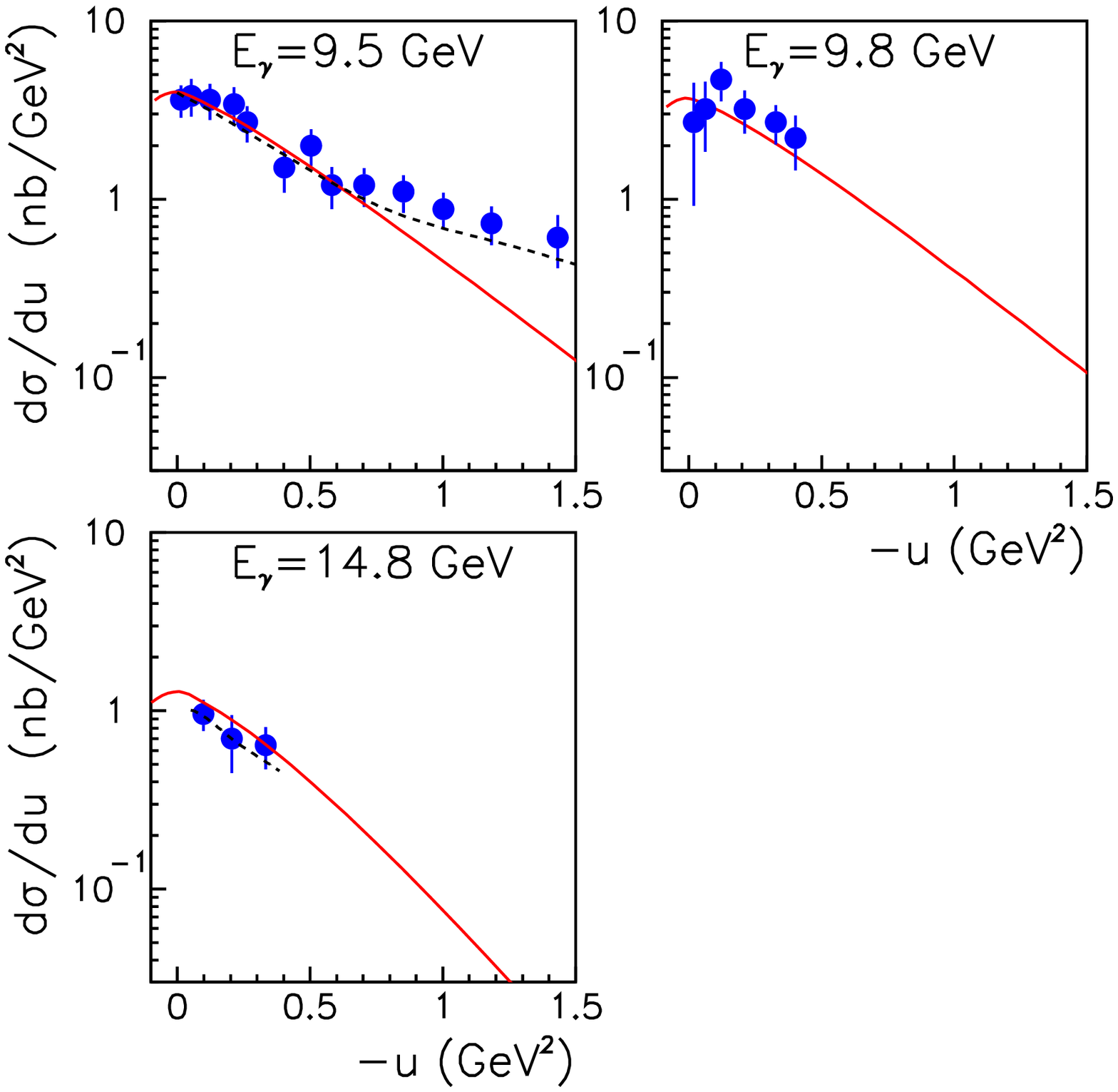,width=9.cm}}
\vspace*{-6mm}
\caption{Differential cross section for $\gamma{p}{\to}\pi^+{n}$ as a
function of the four-momentum transfer squared $u$ at different photon
energies $E_\gamma$. The circles are data from
Refs.~\cite{Anderson1,Anderson0}. The solid lines are the results of our
Regge model. The results of Ref.~\cite{Storrow1} (dashed lines) are
included at the energies where they are available. 
}
\label{phob3}
\end{figure}

Unfortunately, there are not that many data points available at energies
above $\sqrt{s}{\simeq}$3 GeV and for backward angles. Some
experimental information on the $\gamma{p}{\to}\pi^0{p}$ differential cross
section obtained at SLAC as a function of $u$ is provided in
Ref.~\cite{Tompkins}. Those data points are shown in Fig.~\ref{phob1}. 
The SLAC measurements on the $\gamma{p}{\to}\pi^+{n}$ differential cross 
section as a function of
the squared four-momentum $u$ were published in
Refs.~\cite{Anderson1,Anderson0}. They are presented in Fig.~\ref{phob3}.
All data points in the range $-u{<}1\,$GeV$^2$ were included in the fit.

\begin{table}[b]
\begin{center}
\caption{Parameters of the model. The ratio of $N_S/N_V$ is -0.08$\pm$0.02.}
\label{tabp}
\begin{tabular}{|c|c|c|c|}
\hline
  & $\gamma_0$  & $\gamma_3$ & $\alpha_0$\\ 
\hline
$N_\alpha$      &72.2$\pm$5.9  &$-$11.6$\pm$3.1   &$-$0.30$\pm$0.09  \\
$N_\gamma$      &0.54$\pm$2.9   & $-$11.6$\pm$0.4   & $-$0.40$\pm$0.07 \\
$\Delta_\delta$ &$-$4.7$\pm$0.9 & 4.0$\pm$0.2   & 0.25$\pm$0.02  \\
\hline
\end{tabular}
\end{center}
\end{table}

The solid lines in Figs.~\ref{phob1} and \ref{phob3} represent the results
of the fit obtained with the parameters listed in Table~\ref{tabp}.
Although, in principle, one could use the intercepts of the baryon
trajectories extracted from the baryon spectra, we treated them as free
parameters. Furthermore, since it turned out that the data are insensitive
to the $\Delta_\beta$-contribution we did not include this trajectory. 
The overall $\chi^2$ achieved amounts to $\chi^2/{\rm ndf} = 1.2$. It is
interesting that the $N_\alpha$, $N_\gamma$ and $\Delta_\delta$
trajectories determined in the fit to the scattering data imply intercept
parameters that are compatible with those extracted from the baryon
spectra, cf. Tables~\ref{summary1} and \ref{tabp}. But the partly
significant uncertainties of the intercepts obtained in the fit, listed in
Table~\ref{tabp} too, indicate that their values are not so well
constrained by the backward-angle scattering data. 

The overall description of the data by our Regge model is of comparable quality 
to those reported in earlier studies \cite{Berger,Storrow1,Beaupre,Barger1}.
For the ease of comparison, we include here the results of the (latest) previous 
analysis of backward pion photoproduction by Storrow and 
Triantafillopoulos~\cite{Storrow1},  
cf. the dashed lines in Figs.~\ref{phob1} and \ref{phob3}\footnote{We 
noticed some discrepancies between the data as given in the Durham data 
base~\cite{Durham}, which we use,
and the data points drawn in the figures of Ref.~\cite{Storrow1},
especially at photon energy of 5 GeV. 
}. 

\begin{figure}[t]
\vspace*{-5mm}
\centerline{\hspace*{2mm}\psfig{file=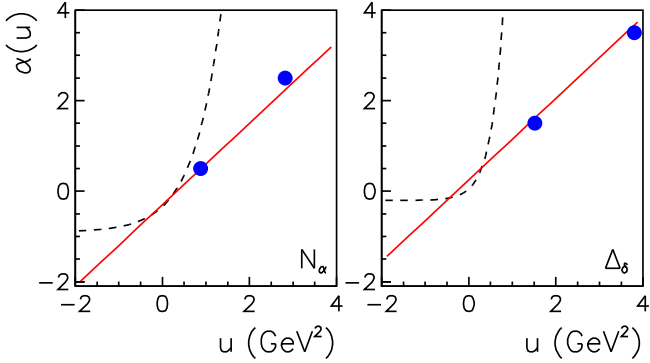,width=10.cm}}
\vspace*{-44mm}
\caption{The trajectories for $N_\alpha$ and $\Delta_\delta$ exchanges
shown as a function of $u$. The solid lines are our parameterizations
given by Eq.~(\ref{traj1}). The dashed lines indicate the trajectories 
used in Ref.~\cite{Storrow1}. The circles show the baryon states 
with relevant quantum numbers.}
\label{phob12}
\end{figure}

The most striking difference in the results is that the model 
of \cite{Storrow1} reproduces the data for larger $-u$ values, 
i.e. also for $-u{>}$1 GeV$^2$, at higher energies. 
This is due to a principal 
difference between our approach and the one of Ref.~\cite{Storrow1}
in the parameterization of the exchange trajectories $\alpha(u)$ 
for the baryon exchanges. While we use linear Regge trajectories (given by
Eq.~(\ref{traj1}) and shown in Fig.~\ref{phob12} by the solid 
lines for the $N_\alpha$ and $\Delta_\delta$ exchanges) 
the trajectories adopted in Ref.~\cite{Storrow1} are nonlinear functions 
of the four-momentum transfer squared $u$, as is illustrated by the 
dashed lines in Fig.~\ref{phob12}. 
Indeed, these nonlinear trajectories, which are modelled by the behaviour
of the data, allow them to describe the experiments for $-u{>}$1 GeV$^2$, 
but they are no longer linked to the baryon spectra
at $u{>}0$, cf. Fig.~\ref{phob12}. 
Interestingly, in the region of small $|u|$, where the Regge approach 
is expected to work best, our results are better in line 
with the features shown by the data, specifically for the reaction
$\gamma{p}{\to}\pi^0{p}$, cf. Fig.~\ref{phob1}.
For completeness, let us mention that the authors of Ref.~\cite{Storrow1} 
also assumed that the $N_\alpha$ and $N_\gamma$ trajectories are the same. 

In other previous works \cite{Berger} several extensions of the Regge
pole model were considered, for example, including absorptive 
corrections. This again allowed the authors to achieve 
a description of the data for larger $-u$ values, however, at the 
expense of practically doubling the number of free parameters that need to 
be determined in the fit. In view of the scarse data for $-u{>}1$ GeV$^2$ 
we refrain from following that strategy. 
Note that also some of the model ansatzes considered in \cite{Berger}
have difficulties in reproducing the data at small $|u|$.

\begin{figure}
\vspace*{-6mm}
\centerline{\hspace*{2mm}\psfig{file=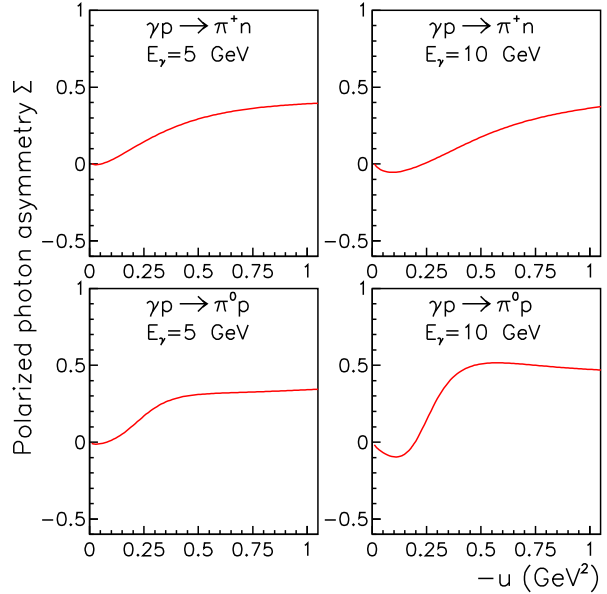,width=9.cm}}
\vspace*{-5mm}
\caption{Polarized photon asymmetry $\Sigma$ for 
$\gamma{p}{\to}\pi^+{n}$ and $\gamma{p}{\to}\pi^0{p}$ 
as a function of $-u$, for different photon energies $E_\gamma$. 
The lines are the results of our Regge model.}
\label{phob11a}
\end{figure}

Predictions for the polarized photon asymmetry $\Sigma$ are shown
in Fig.~\ref{phob11a} for two energies. 

\begin{figure}
\vspace*{-5.mm}
\centerline{\hspace*{4mm}\psfig{file=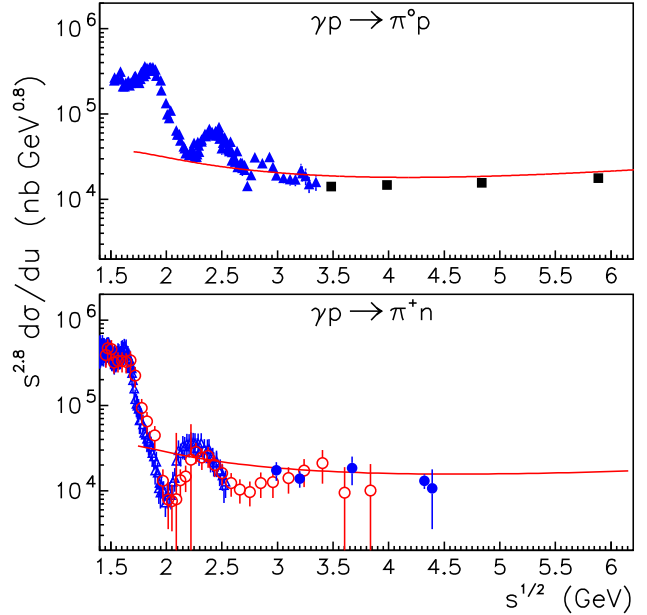,width=9.4cm}}
\vspace*{-4mm}
\caption{Differential cross sections for $\pi^0$ (upper panel) and $\pi^+$
(lower panel) photoproduction at $\vartheta{=}180^o$ as a function of the
invariant collision energy $\sqrt{s}$. Data for the $\pi^0p$ channel are
from Refs.~\cite{Buschorn} (triangles) and \cite{Tompkins} (squares).
Data for the $\pi^+n$ channel are from Refs.~\cite{Ekstrand} (open circles),
\cite{Bouquet} (triangles) and ~\cite{Anderson1,Anderson0} (filled circles).
The lines are the results of our Regge model. Note that all results are 
multiplied with a factor $s^{2.8}$, cf. text.}
\label{phob2}
\end{figure}

\section{Pion photoproduction at {\boldmath$180^\circ$}}
The experimental information on $\pi^0$- and $\pi^+$-meson photoproduction 
at $\vartheta{=}180^\circ$ is shown in Fig.~\ref{phob2}  
for a large range of photon energies. 
Note that the data from Refs.~\cite{Buschorn} (filled triangles), 
\cite{Ekstrand} (open circles) and \cite{Bouquet} (open triangles)
were not included in our fit, since only few points are available at 
high energies. The lines in
Fig.~\ref{phob2} are the results of our model calculation. We have
multiplied the data and also the predictions of our model with a factor
$s^{2.8}$ in order to facilitate the comparison between data and theory.
This factor corresponds to the energy dependence of the leading Regge
trajectory. For energies from around 3~GeV upwards the data seem to
approach the high-energy limit inferred from the Regge fit. 

\begin{figure}[t]
\vspace*{-2.7mm}
\centerline{\hspace*{4mm}\psfig{file=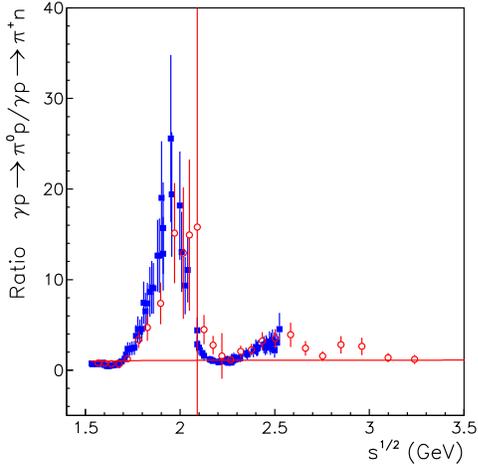,width=7cm}}
\vspace*{-3mm}
\caption{Ratio of the $\pi^0$ to $\pi^+$ photoproduction cross sections at
$\vartheta{=}180^o$ as a function of the invariant collision energy. The
filled squares are the results obtained with the data from
Refs.~\cite{Buschorn,Bouquet} while open circles show the ratios obtained
with the data from Refs.~\cite{Buschorn,Ekstrand}. The line is the ratio
from our Regge model.}
\label{phob7a}
\end{figure}

At invariant energies from 2 to 3~GeV the data exhibit oscillations around
the continuation of the Regge result.
It is interesting to observe that the differential cross sections for
$\gamma{p}{\to}\pi^0p$ and $\gamma{p}{\to}\pi^+n$ at
$\vartheta{=}180^\circ$ show a very different energy dependence. This is
best seen in Fig.~\ref{phob7a} where we present the ratio of the $\pi^0$ to
$\pi^+$ cross sections at $\vartheta{=}180^o$. The line is the ratio
obtained for the Regge result which is close to 1.12. The ratio of the
experimental values varies strongly with energy and indicates the presence
of structures around $\sqrt{s}{\simeq}$2 and 2.5~GeV. There could be also a
structure around $\sqrt{s}{\simeq}\,$2.9 GeV but here the accuracy of the
data is not sufficient for drawing reliable conclusions. 

\section{Photoproduction at large angles}

\begin{figure}[t]
\vspace*{-6mm}
\centerline{\hspace*{2mm}\psfig{file=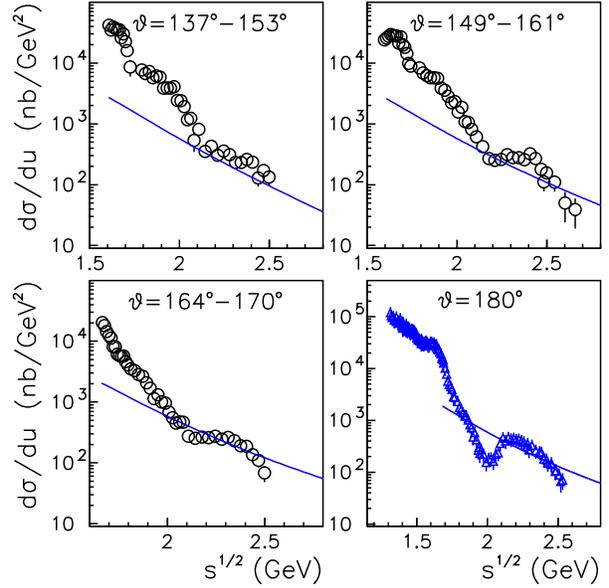,width=9.cm}}
\vspace*{-5mm}
\caption{Differential cross sections for $\gamma{p}{\to}\pi^+{n}$ as a
function of the invariant collision energy $\sqrt{s}$, for different
production angles $\vartheta$ in the cm system. The open circles are data
from Ref.~\cite{Alvarez}, while the open triangles are from
Ref.~\cite{Bouquet}. The lines are the results of our Regge model. }
\label{phob2e}
\end{figure}

Differential cross sections for $\gamma{p}{\to}\pi^+{n}$ for several
angular ranges in the region $\vartheta{>}135^0$ are shown in
Fig.~\ref{phob2e} as a function of the invariant collision energy. The data
are taken from Refs.~\cite{Bouquet,Alvarez}. The lines indicate the results
of our Regge model for $\vartheta=145^\circ$, $155^\circ$, $167^\circ$, and
$180^\circ$, respectively. 
For invariant energies below 2~GeV the data lie above the
high-energy limit as given by the Regge calculation. For higher energies
the data vary around the Regge predictions and those variations become more
pronounced with increasing photoproduction angle. It is interesting to note
that at the lowest photoproduction angle considered the data seem to
approach the high-energy limit -- and even at fairly low energies. 

\begin{figure}
\vspace*{-5.5mm}
\centerline{\hspace*{2mm}\psfig{file=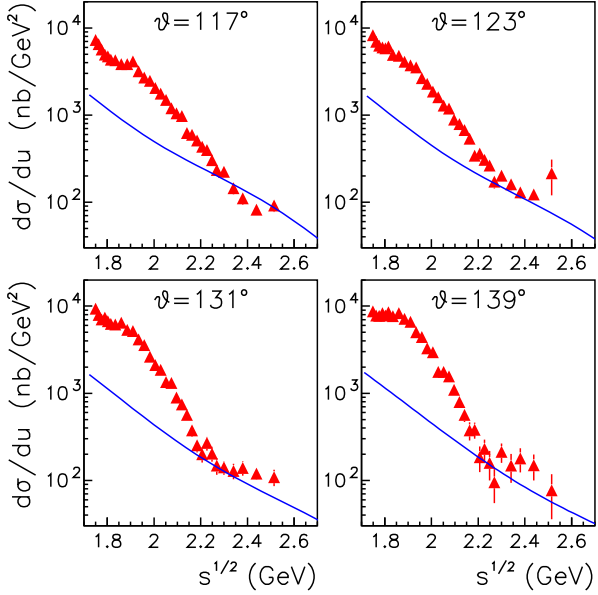,width=9.cm}}
\vspace*{-4mm}
\caption{Differential cross sections for $\gamma{p}{\to}\pi^0{p}$ as a
function of invariant collision energy $\sqrt{s}$, for different production
angles $\vartheta$ in the cm system. The data are from the ELSA
collaboration~\cite{Bartholomy}. The lines are the results of our Regge
model. }
\label{phob2c}
\end{figure}

Data on neutral pion photoproduction at large angles were taken at
ELSA~\cite{Bartholomy}, JLab~\cite{Dugger} and LEPS~\cite{Sumihama}.
Fig.~\ref{phob2c} contains the data from ELSA. The lines are our Regge
results for the $\gamma{p}{\to}\pi^0{p}$ reaction. Here the data lie above
the calculations up to somewhat higher energies, i.e. up to 2.2-2.3~GeV. It
seems that the data approach the high-energy limit in this energy region
but reliable conclusions are not possible because of the limited accuracy
of the data. 

\begin{figure}[t]
\vspace*{-5.5mm}
\centerline{\hspace*{2mm}\psfig{file=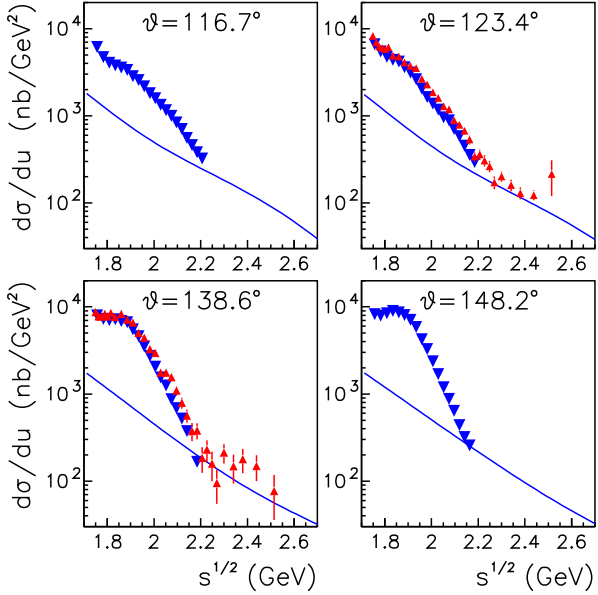,width=9.cm}}
\vspace*{-4mm}
\caption{Differential cross section for $\gamma{p}{\to}\pi^0{p}$ as a
function of the invariant collision energy $\sqrt{s}$, for different
production angles $\vartheta$ in the cm system. The inverted triangles are
data from JLab~\cite{Dugger}, while triangles are experimental results from
ELSA~\cite{Bartholomy}. The lines are the results of our Regge model.}
\label{phob2d}
\end{figure}

\begin{figure}
\vspace*{-6mm}
\centerline{\hspace*{2mm}\psfig{file=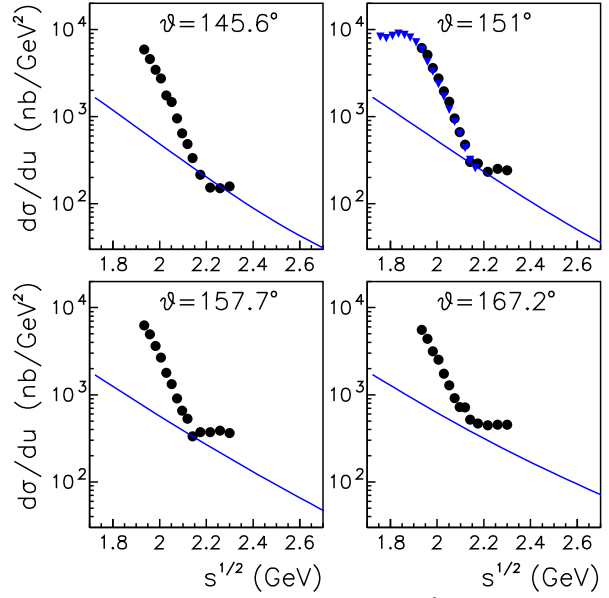,width=9.cm}}
\vspace*{-4mm}
\caption{Differential cross section for $\gamma{p}{\to}\pi^0{p}$ as a function
of the invariant collision energy $\sqrt{s}$, for different production
angles $\vartheta$ in the cm system. The solid circles are data from 
LEPS~\cite{Sumihama}, while the filled inverted triangles are results from
JLab~\cite{Dugger}. The lines are the results of our Regge model.}
\label{phob2b}
\end{figure}

\begin{figure}
\vspace*{-6mm}
\centerline{\hspace*{2mm}\psfig{file=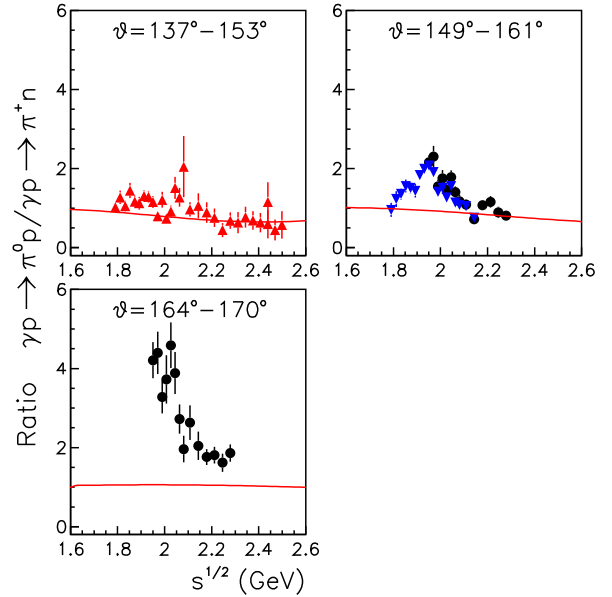,width=9cm}}
\vspace*{-4mm}
\caption{Ratio of the $\pi^0$ to $\pi^+$ photoproduction cross sections at
different angles $\vartheta$ as a function of the invariant collision
energy, based on the $\pi^+$ photoproduction data of \cite{Alvarez}.
The triangles are the ratio for the $\pi^0$ data from
ELSA~\cite{Bartholomy}, the circles for the LEPS~\cite{Sumihama} data, and
the inverted triangles for the JLab~\cite{Dugger} data. The line is the
ratio from our Regge model. }
\label{phob7b}
\end{figure}

The recent Jefferson Lab data \cite{Dugger} on $\pi^0$-meson
photoproduction are shown by inverted triangles in Fig.~\ref{phob2d}
together with the Regge result. Note that in the backward region the JLab
measurement extends only up to around $\sqrt{s} \simeq 2.2$~GeV. At the
angles $\vartheta{=}123^0$ and $\vartheta{=}139^0$ we can compare the data
with corresponding experimental information from ELSA~\cite{Bartholomy}
(triangles). Obviously, in the kinematical range relevant for our study the
two measurements are in reasonable agreement with each other. 
At last, in Fig.~\ref{phob2b} we present the data for
$\gamma{p}{\to}\pi^0{p}$ from SPring-8 at LEPS~\cite{Sumihama} (circles).
Also here a comparison with the JLab results~\cite{Dugger} is possible at a
particular angle ($\vartheta{=}148.2^0$), cf. the inverted triangles. These
measurements agree nicely with each other, too. 
 
Evidently, the LEPS results do not approach the high-energy limit as
inferred from the Regge fit to high-energy data but rather tend to deviate
more strongly from the predictions with increasing energy. Thus, at least
within the energy region of $\sqrt{s}{\le}2.5$~GeV, covered by the new
measurements of $\gamma{p}{\to}\pi^0{p}$ differential cross sections at
large angles~\cite{Bartholomy,Dugger,Sumihama}, there is no clear signal
for  convergence. Therefore, it remains an open question from which
energies onwards Regge phenomenology might be applicable. Though the old
data~\cite{Buschorn,Ekstrand,Bouquet} for $\vartheta{=}180^\circ$ that
extend to higher energies provide some information with regard to this
issue, their precision is not sufficient for drawing reliable conclusions. 

The $\pi^0$ and $\pi^+$-meson photoproduction cross sections at large
angles differ significantly as was already illustrated by the
$\pi^0{/}\pi^+$ cross-section ratio at $\vartheta{=}180^\circ$ in the
Fig.~\ref{phob7a}. Corresponding ratios for other angles in the backward
region are presented in Fig.~\ref{phob7b}. It is interesting to see that at
those angles the ratio is much smaller. Specifically, in the range
$137^\circ - 153^\circ$ it is even smaller than 2 and, moreover,
practically energy independent. 

\begin{figure}
\vspace*{-6mm}
\centerline{\hspace*{2mm}\psfig{file=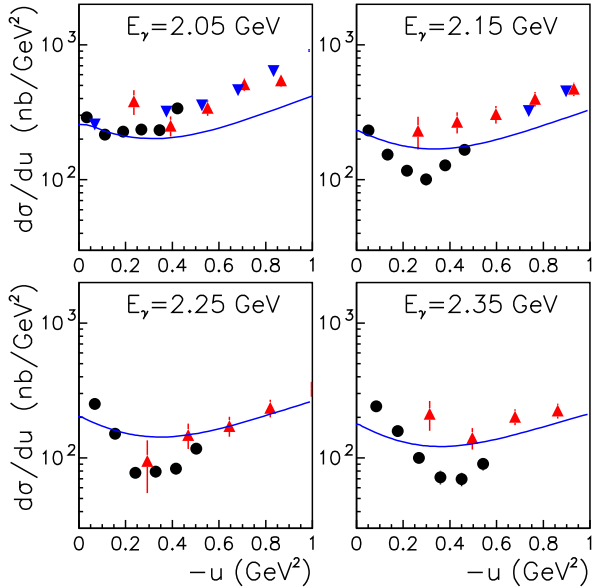,width=9.cm}}
\vspace*{-5mm}
\caption{Differential cross section for $\gamma{p}{\to}\pi^0{p}$ as a
function of $-u$, for different photon energies $E_\gamma$. The energies
indicated in the figures correspond to the invariant collision energies
$\sqrt{s}{\simeq}$2.17, 2.22, 2.26 and 2.3~GeV, respectively. 
The triangles are data from ELSA~\cite{Bartholomy}, the circles are data
from LEPS~\cite{Sumihama}, and the inverted triangles are results from
JLab~\cite{Dugger}. The lines are the results of our Regge model. }
\label{phob10}
\end{figure}

Finally, we compare the $u$ dependence of the data from ELSA, JLab and LEPS 
at the highest measured energies. 
The differential cross sections 
from the different experiments are shown in Fig.~\ref{phob10} in 
the range $-1 \le u \le 0$. 
The data presented in the figure were taken at almost the same photon energies, 
corresponding to invariant collision energies in the range from 2.17 to 2.3~GeV. 
Fig.~\ref{phob10} illustrates that the LEPS data are partly in 
strong disagreement with the ELSA measurement. At the two highest energies
the Regge calculation reproduces the ELSA data reasonably well. 
The LEPS data exhibit a quite different $u$-de\-pend\-ence as compared to
the Regge result and they differ significantly in the absolute value too. 
In any case, one should keep in mind that 
at such low energies the accessible range is small for 
$u$ as well as $t$ so that the regions of small $|u|$ and of
small $|t|$ tend to overlap. Therefore, one expects that Regge contributions 
from both channels should come into play and interferences will occur. 
But in our model only $u$ channel poles are taken into account. 
Moreover, there are well-established (four-star) $N$- and $\Delta$
resonances in the mass region of 2.19 to 2.42~GeV \cite{PDG} which
should have an impact on the cross sections. 
Thus, our predictions shown in Fig.~\ref{phob10} have primarily an 
illustrative character. 

\section{Summary}

We have performed a systematic analysis of backward photoproduction of
pions in the $\gamma{p}{\to}\pi^0{p}$ and $\gamma{p}{\to}\pi^+{n}$
reactions. Regge phenomenology was applied at invariant collision energies
above 3~GeV in order to fix the reaction amplitude. The aim of our study
was to see whether we can find any clues regarding the energy region where
the data approach the high-energy limit as given by the Regge calculation.

The data~\cite{Buschorn} on $\pi^0$-meson photoproduction at
$\vartheta{=}180^\circ$ indicate that this limit could be reached possibly
at energies of around $\sqrt{s}\simeq 3$~GeV. The most recent results on
neutral pion photoproduction at large angles available from
ELSA~\cite{Bartholomy}, JLab~\cite{Dugger} and LEPS~\cite{Sumihama} cover
energies up to $\sqrt{s}$=2.5 GeV. Within this energy region there is no
clear signal for a convergence towards the results inferred from our Regge
fit to high-energy data. 
Experiments~\cite{Ekstrand,Bouquet} on $\pi^+$-meson photoproduction at
$\vartheta{=}180^\circ$ suggest that in this channel the data might
approach the high-energy limit likewise at roughly $\sqrt{s}{\simeq}$3~GeV.
However, as in case of $\pi^0$ photoproduction, data with higher precision
would be needed to allow for a more solid statement.
 
It is difficult to say whether the new
measurements~\cite{Bartholomy,Dugger,Sumihama} indicate any prominent
features that could be due to the excitation of high-mass baryons. One
expects that backward scattering is the best angular region to find signals
from the excitation of baryons. Indeed the experimental
results~\cite{Barger9,Crittenden} available for $\pi{p}$ scattering at
backward angles indicate directly a strong variation of the differential
cross section up to invariant energies of $\sqrt{s}{\simeq}2.9$~GeV.
Unfortunately, the new backward photoproduction data do not show any such
clean features.

\begin{acknowledgement}
We would like to thank T. Nakano and M. Sumihama for sending 
us the LEPS data. We appreciate discussions with H.~Gao, M.~Dugger, 
E.~Klempt, E.~Pasyuk, U.~Thoma.
This work is partially supported by the Helmholtz Association through funds
provided to the virtual institute ``Spin and strong QCD'' (VH-VI-231), by \,
the European Community-Research
Infrastructure Integrating Activity ``Study of Strongly Interacting
Matter'' (acronym HadronPhysics2, Grant Agreement no. 227431) under the
Seventh Framework Programme of EU, and by DFG (SFB/TR 16, ``Subnuclear
Structure of Matter'').
F. H. is grateful for the support from the Alexander von Humboldt Foundation.
A.S. acknowledges support by the JLab grant SURA-06-C0452 and the
COSY FFE grant No. 41760632 (COSY-085).
\end{acknowledgement}


\begin{thebibliography}{99}
\bibitem{Bartholomy}
	O. Bartholomy {\it et al.}, Phys. Rev. Lett.{\bf 94}, 012003 (2005);
	H.~van Pee {\it et al.}, 
  Eur.\ Phys.\ J.\  A {\bf 31}, 61 (2007) [arXiv:0704.1776].
\bibitem{Dugger}
	M. Dugger {\it et al.}, Phys. Rev. C {\bf 76}, 025211 (2007)
	[arXiv:0705.0816].
\bibitem{Sumihama}
	M. Sumihama {\it et al.}, Phys. Lett. B {\bf 657}, 32 (2007)
	[arXiv:0708.1600].
\bibitem{PDG}
	C. Amsler {\it et al.}, Phys. Lett. B {\bf 667}, 1 (2008).
\bibitem{Arndt}
	R. Arndt, W.J. Briscoe, I.I. Strakovsky and R.L. Workman,
	Phys. Rev. C {\bf 66}, 055213 (2002);
        \hbox{http://gwdac.phys.gwu.edu}. 
\bibitem{Arndt2}
	R. Arndt, W. Briscoe, I. Strakovsky and R. Workman,
	Eur. Phys. J. A {\bf 35}, 311 (2008).
\bibitem{Berger2}
	E.L. Berger and G.C. Fox, Nucl. Phys. B {\bf 30}, 1 (1971).
\bibitem{Hoehler}
	G. H{\"o}hler, Landolt-B{\"o}rnstein {\bf 9}, Springer Verlag, 
	Berlin, 1983.	
\bibitem{Irving}
	A. C. Irving and R. P. Worden, Phys. Rept. {\bf 34}, 117 (1977).	
\bibitem{Huang}
	F. Huang, A. Sibirtsev, S. Krewald, C. Hanhart, J. Haidenbauer
	and Ulf-G. Mei{\ss}ner, arXiv:0810.2680, Eur. Phys. J. A, 
        in print. 
\bibitem{Barger9}
	V. Barger and D. Cline, Phys. Rev. {\bf 155}, 1792 (1967).
\bibitem{Crittenden}
	R.R. Crittenden, R.M. Heinz, D.B. Lichtenberg and E. Predazzi,
	Phys. Rev. D {\bf 1}, 169 (1970).
\bibitem{Berger}
	E.L. Berger and G.C. Fox, Nucl. Phys. B {\bf 26}, 1 (1970).
\bibitem{Sibirtsev1}	
	A. Sibirtsev, J. Haidenbauer, S. Krewald, T.S.H. Lee, Ulf-G.
	Mei{\ss}ner and A.W. Thomas, Eur. Phys. J. A {\bf 34}, 49 (2007)
        [arXiv:0706.0183].
\bibitem{Jacob}
	M. Jacob and G.C. Wick, Ann. Phys. {\bf 7}, 404 (1959).
\bibitem{Gell-Mann}
 	M. Gell-Mann, M.L. Goldberger, F.E. Low, E. Marx and F.~Zachariasan,
 	Phys. Rev. {\bf 133}, B145 (1964).	
\bibitem{Storrow2}
	J.K Storrow, Electromagnetic Interactions of Hadrons {\bf 1}, 
	edited by A. Donnachie and G. Shaw, Plenum, NY (1978), pp. 263. 
\bibitem{Chew}
 	G.F. Chew, M.L. Goldberger, F.E. Low and Y. Nambu, Phys. Rev. {\bf
	106}, 1345 (1957).	
\bibitem{Ball}
	J.S. Ball, Phys. Rev. {\bf 124}, 2014 (1961).	
\bibitem{Storrow1}
	J.K Storrow and E. Triantafillopoulos, J. Phys. G {\bf 4}, 1679 (1978).
\bibitem{Paschos}
	E.A. Paschos, Phys. Rev. Lett. {\bf 21}, 1855 (1968).
\bibitem{Beaupre}
	J.V. Beaupre and E.A. Paschos, Phys. Rev. D {\bf 1}, 2040 (1970).
\bibitem{Barger1}
	V. Barger and P. Weiler, Nucl. Phys. B {\bf 20}, 615 (1970).	
\bibitem{MacDowell}
	S.W. MacDowell, Phys. Rev. {\bf 116}, 774 (1959).
\bibitem{Gribov}
	V.N. Gribov, JETP {\bf 16}, 1080 (1963).
\bibitem{Jackson}
	J.D. Jackson and G.E. Hite, Phys. Rev. {\bf 169}, 1248 (1968).
\bibitem{Barker}
	I.S. Barker, A. Donnachie and J.K. Storrow, Nucl. Phys. B 
	{\bf 95}, 347 (1975). 
\bibitem{Benmerrouche}
	M. Benmerrouche, N.C. Mukhopadhyay and J.F. Zhang, 
	Phys. Rev. D {\bf 51}, 3237 (1995).
\bibitem{Barger4}
	V. Barger and D. Cline, Phys. Rev. Lett. {\bf 20}, 298 (1968).
\bibitem{Fiore}
	R. Fiore, L.L. Jenkovszky, F. Paccanoni and A. Prokudin, Phys. Rev.
	D {\bf 70}, 054003 (2004).
\bibitem{Minkowski}
	P.~Minkowski,
	Lett.\ Nuovo Cim.\  {\bf 3S1}, 503 (1970)
	[Lett.\ Nuovo Cim.\  {\bf 3}, 503 (1970)].
\bibitem{Carlitz}	
	R. Carlitz and M. Kislinger, Phys. Rev. Lett. {\bf 24}, 186 (1970).
\bibitem{Barger3}
	V. Barger and D. Cline, Phenomenological Theories of High Energy
	Scattering,  New York, Benjamin (1969).
\bibitem{Storrow3}
	J.K. Storrow and G.A. Winbow, Nucl. Phys. B {\bf 54}, 560 (1973).
\bibitem{Winbow}	
	J.K. Storrow and G.A. Winbow, J. Phys. G  {\bf 1}, 263 (1975).
\bibitem{Tompkins}
	D.H. Tompkins {\it et al.}, Phys. Rev. Lett. {\bf 23}, 725 (1969). 
\bibitem{Anderson1}
	R. Anderson  {\it et al.}, Phys. Rev. Lett. {\bf 21}, 479 (1968). 
\bibitem{Anderson0}
	R. Anderson  {\it et al.}, Phys. Rev. D {\bf 14}, 679 (1976). 
\bibitem{Durham}
	The Durham HEP Databases,  http://durpdg.dur.ac.uk/.
\bibitem{Buschorn}
	G. Buschhorn {\it et al.}, Phys. Rev. Lett. {\bf 20}, 230 (1968).
\bibitem{Ekstrand}
	K. Ekstrand {\it et al.}, Phys. Rev. D {\bf 6}, 1 (1972).
\bibitem{Bouquet} 
	B. Bouquet {\it et al.}, Phys. Rev. Lett. {\bf 27}, 1244 (1971).
\bibitem{Zhu1}
	L.Y. Zhu {\it et al.}, Phys. Rev. C {\bf 71}, 044603 (2005)  
	[arXiv:nucl-ex/0409018].
\bibitem{Zhu2}
	L.Y. Zhu {\it et al.}, Phys. Rev. Lett. {\bf 91}, 0222003 (2003)  
	[arXiv:nucl-ex/0211009].
\bibitem{Alvarez}
	R.A. Alvarez {\it et al.}, Phys. Rev. D {\bf 1}, 1946 (1970).
\bibitem{Bussey}
	P.J. Bussey et al., Nucl. Phys. B {\bf 154}, 492 (1979).
\end{thebibliography}
\end{document}